\documentstyle[preprint,prl,aps]{revtex}
\begin{document}

\draft

\title{Competition Between Antiferromagnetic Order
and Spin-Liquid Behavior in the Two-Dimensional Periodic
Anderson Model at Half-Filling}

\author{M. Veki\'c,$^{1}$ J.W. Cannon,$^{2}$ D.J. Scalapino,$^{3}$
R.T. Scalettar,$^{1}$ and R.L. Sugar$^{3}$}
\address{$^{1}$Department of Physics, University of California,
Davis, CA 95616}
\address{$^{2}$Physics Department, Centenary College, 2911 Centenary
Blvd., Shreveport, LA 71104}
\address{$^{3}$Department of Physics, University of California,
Santa Barbara, CA 93106}

\date{\today}

\maketitle

\begin{abstract}
We study the two-dimensional periodic Anderson model at half-filling
using quantum Monte Carlo (QMC) techniques.
The ground state undergoes a magnetic order-disorder
transition as a function of the effective exchange coupling between the
conduction and localized bands. Low-lying spin and
charge excitations are determined using the maximum entropy
method to analytically continue the QMC data. At finite temperature we find
a competition between the Kondo effect and antiferromagnetic order which
develops in the localized band through Ruderman-Kittel-Kasuya-Yosida
interactions.
\end{abstract}



The periodic Anderson model (PAM) \cite{anderson} describes
a localized $f$-band of strongly correlated electrons
hybridized with a $d$-band of free conduction
electrons. As the temperature, filling, and Hamiltonian parameters
are varied, there is a competition between the
Ruderman-Kittel-Kasuya-Yosida (RKKY)
interaction \cite{rkky} and the Kondo effect \cite{kondo}.
The RKKY interaction favors ordering
the magnetic moments
of the localized $f$-band, while the Kondo effect screens the localized
magnetic moments and quenches
the magnetic interaction through the formation of singlets between the two
bands \cite{review}. The Kondo effect is expected to dominate for
large exchange coupling between the conduction electrons and the local
moments. For smaller values of the exchange interaction, instead, the local
moments order, in most cases antiferromagnetically.

It is believed that this behavior qualitatively describes the competition
between magnetic ordering and singlet formation in a number of the
heavy fermion materials.  Additionally,
several small-gap semiconducting compounds involving either rare-earth or
actinide metals \cite{experiments} can be well described by the PAM. These
systems exhibit local moment behavior at high temperatures. At low
temperatures the hybridization of the local moments with the
conduction band leads to magnetic ordering and the formation of a small
energy gap or, in some cases, to a Kondo insulator with no long-range
magnetic order.

The competition between the RKKY interaction and the Kondo effect has
been studied within the framework of the two-impurity Anderson
model \cite{fyescalapino,coleman,twoimpurities,sakai,oneimpurity},
the one-dimensional PAM \cite{onedim} and the Kondo lattice
\cite{kondolattice},
and in infinite dimensions \cite{infinited}.
Here we present QMC results for the two-dimensional PAM
at half-filling. This is an interesting model since it exhibits various
types of insulating states ranging from an antiferromagnetic insulator
to a Kondo insulator or, if the $f$-$d$ hybridization dominates, a
simple band-insulator.

The Hamiltonian for the two-dimensional PAM is
\begin{eqnarray}
\hat H & = & -t\sum_{\langle i,j\rangle\sigma} \left(
d^{\dagger}_{i\sigma}
d_{j\sigma}+d^{\dagger}_{j\sigma}d_{i\sigma}\right)
+\epsilon_d\sum_{i\sigma} n^d_{i\sigma}
+ U_{f}\sum_i \left( n^f_{i\uparrow}-{1\over 2}\right)
\left( n^f_{i\downarrow}-{1\over 2}\right) \cr & & +
\epsilon_f\sum_{i\sigma} n^f_{i\sigma} \label{hamiltonian}
- V\sum_{i\sigma} \left(
d^{\dagger}_{i\sigma}f_{i\sigma}+f^{\dagger}_{i\sigma}d_{i\sigma}\right).
\end{eqnarray}
Here $t$ is the hopping parameter in the $d$-band, $V$ is the
hybridization energy between the bands, $U_{f}$ is the Coulomb repulsion
in the $f$-band, $\epsilon_f$ and $\epsilon_d$ are the energy levels of
the $d$- and $f$-band respectively, and
$n^d_{i\sigma}\equiv d^\dagger_{i\sigma}d_{i\sigma}$ and
$n^f_{i\sigma}\equiv f^\dagger_{i\sigma}f_{i\sigma}$
are the density operators for the two bands. In the following
we will set $t=1$ and consider the symmetric case
($\epsilon_f=0$) at half-filling ($\epsilon_d=0$).

The zero temperature behavior of the PAM along the $U_{f}=0$ and $V=0$
axes in the $U_f$-$V$ phase diagram is simple.  In the former case,
we have a ``band-insulator''
as the hybridization $V$ opens up a gap where the dispersionless
$f$ electron level crosses the conduction band.
In the latter case we have a set of individual local moments
completely decoupled from the conduction band.
When $V$ and $U_{f}$ are both non-zero,
a qualitative understanding of the
zero temperature order-disorder transition can be obtained from
a comparison of the RKKY energy scale $J_{\mbox{\tiny RKKY}} \sim J^{2}/W$
and the Kondo energy scale $T_K \sim We^{-W/J}$, where $J\sim V^{2}/U_{f}$ and
$W$ is the bandwidth.
For small values of $J$, long-range RKKY driven antiferromagnetic order will
set in. As the exchange coupling $J$ is increased beyond a fixed value,
we expect that the order will be quickly suppressed due to the
formation of spin singlets between the two bands.
The above argument suggests a
functional form of the phase boundary with $U_f \propto V^{2}$.

In order to determine whether the ground state of the PAM does have
long-range antiferromagnetic order, we have carried out a finite size scaling
analysis of the equal time spin-spin correlations obtained with QMC
to calculate the staggered $f$-orbital magnetization $m_f$ in the
thermodynamic limit.
When this quantity is non-zero, the ground state of the PAM has long-range
antiferromagnetic order. In addition, we have evaluated non-equal time
correlations to extract the dynamic
$f$-orbital spin and charge susceptibilities as well as the low
temperature dependence of the total uniform spin and charge
susceptibilities in order to determine the spin and charge gaps. For an
antiferromagnetic Mott insulator the charge gap is finite but there are
gapless Goldstone antiferromagnetic spin excitations.
However, in a Kondo insulator there are gaps for all
excitations (with the charge gap larger than the spin gap)
and consequently no long-range order. In the PAM we expect the
small exchange interaction phase to be well described by a Mott insulator
with antiferromagnetic order,
while the larger exchange interaction phase should be well
described by a Kondo insulator with spin-liquid behavior.
In addition, in the limit of a large exchange interaction there is a cross-over
to a band-insulator regime where the spin and charge gaps become identical.

At finite temperature we find two different behaviors depending on whether
the parameters are such that the ground state is in the order or disordered
part of the phase diagram. For small
values of the exchange coupling we find a metallic behavior, with
a Kondo resonance developing in the
$f$-band density of states for temperatures below the Kondo temperature
$T_K$. This resonance is then suppressed at a lower temperature, leading
to insulating behavior, by the formation of a gap when the antiferromagnetic
correlations reach the size of the system. The staggered spin
susceptibility tends to diverge at the same temperature at which the charge gap
forms in $N^f(\omega)$.
The uniform spin susceptibility displays a maximum below $T_K$ and tends to
saturate at a finite value as $T$ goes to zero. On the other hand, the uniform
charge susceptibility also displays a maximum but then decays
quickly to zero, indicating the existence of a charge gap.
For large values of the exchange coupling the $f$-band density of
states shows an insulating gap for all finite temperatures and the absence
of antiferromagnetic order. Both the uniform $q\rightarrow 0$
spin and charge susceptibilities
decay to zero as the temperature is lowered, indicating the presence of a gap.

In order to determine the $U_f$-$V$ boundary
separating the antiferromagnetically
ordered from the disordered phase we have calculated the $f$-band spin
correlation function
\begin{equation}
c^f(l_x,l_y)={1\over N}\sum_i \langle m^f_{i+l} m^f_{i} \rangle,
\label{correlation}
\end{equation}
with $m_i^f=n^f_{i\uparrow}-n^f_{i\downarrow}$,
and the $f$-band antiferromagnetic structure factor
\begin{equation}
S^f(\pi,\pi)=\sum_l (-1)^{l_x+l_y} c^f(l_x,l_y),
\label{structure}
\end{equation}
as functions of temperature and system size $N$.
In particular, for a given lattice size, we found that $S^f(\pi,\pi)$
saturates for large values of $\beta$. We have used the saturated values
along with the spin-wave scaling forms
\begin{equation}
m^2_f = {{3S^f(\pi,\pi)}\over {N}} + O\left( {1\over {\sqrt{N}}} \right),
\label{spinwave1}
\end{equation}
and
\begin{equation}
m^2_f = 3c^f(\sqrt{N}/2,\sqrt{N}/2) + O\left( {1\over {\sqrt{N}}} \right),
\label{spinwave2}
\end{equation}
to extract the staggered magnetization $m^2_f$ of the $f$-band.
Figure~1 shows the results of these calculations for two values of the
hybridization $V=1.0$ and $V=1.2$ with $U_f=4$. The system sizes ranged
from $4\times 4$ to $8\times 8$ lattices. For $V=1.0$, the intercept
of a linear fit to the points gives a finite value for
$m^2_f$ in the infinite system. However, for $V=1.2$, $S^f(\pi,\pi)$
does not appear to scale with system size $N$ and, as shown in Fig.~1,
we conclude that $m_f^2=0$.
Similar conclusions can be reached by
scaling $c^f(\sqrt{N}/2,\sqrt{N}/2)$ according to the spin-wave theory result.
In Fig.~1 we show that the extrapolated value for $m^2_f$ obtained from
Eq.(\ref{spinwave2})
is consistent with the value obtained from $S^f(\pi,\pi)$
for $V=1.0$. Again, for $V=1.2$ it appears that $m_f^2=0$.

We have carried out a similar scaling analysis for several values of
$U_f$ and in Fig.~2 we show the $U_f-V$ phase diagram obtained in this
manner.
The boundary is shown as a solid line passing through the
three points obtained from the QMC data at $U_f=2,4$, and $6$.
For small values of
$J\sim V^2/U_f$ we have carried out a mean-field calculation assuming
that
$\langle n^f_{i\sigma}\rangle = {1\over 2}+\sigma {{m^f}\over 2} (-1)^i$.
The dotted curve in Fig.2 shows where the mean-field $m_f$ vanishes.
The antiferromagnetic region obtained from MFT is significantly larger
than that found from QMC due to the effect of fluctuations which are
neglected in MFT.

In order to extract dynamic properties of the system, we have also calculated
imaginary-time properties. Using a
maximum entropy procedure \cite{maxent} to analytically
continue the QMC data, we have
determined the density of states and
the imaginary part of the spin and charge dynamic susceptibilities.
We have inverted the integral
relation relating the single particle Green's function
$G^f_{ij}(\tau)\equiv -\langle T_\tau c_i(\tau)c^\dagger_j(0)\rangle$
to the density of states $N^f(\omega)$,
\begin{equation}
G^f_{ii}(\tau)=-\int_{-\infty}^\infty d\omega {{N^f(\omega)
e^{-\tau\omega}} \over {1+e^{-\beta\omega}}}.
\label{densitystates}
\end{equation}
In Fig.~3 we show $N^f(\omega)$ for several values of $\beta$ for
$U_f=4$ and $V=0.75$. As the temperature is lowered from $\beta=4$ to
$\beta=12$ we see a Kondo resonance peak developing at the Fermi level,
corresponding to the screening of the local moments by the
conduction electrons. However, for finite values of $V$, as
the temperature is further lowered, we find that the Kondo
peak is suppressed until at $\beta=20$ a well developed gap is formed.
On the same lattice size and for the same range of
temperatures we find that the
antiferromagnetic structure factor grows rapidly before saturating to a
finite value at around $\beta=20$. Thus, we associate the gap with the
long-range antiferromagnetic order which sets in at the same temperature
for this lattice size.
For larger values of $V$ the behavior of $N^f(\omega)$
is quite different,
since we do not find any resonance peak at $\omega=0$ but rather
an insulating gap is always present at all temperatures.
It is found that this gap grows with $V$ for any fixed $U_f$.

We have also analytically continued the imaginary-time $f$-orbital
spin and charge
correlation functions
$\chi^f_{ii}(\tau)\equiv \langle T_\tau m_i^f(\tau)m_i^f(0)\rangle$ and
$\Pi^f_{ii}(\tau)\equiv \langle T_\tau n_i^f(\tau)n_i^f(0)\rangle$
in order to calculate the imaginary part of the dynamic spin and charge
susceptibilities Im$\chi^f(\omega)$ and Im$\Pi^f(\omega)$, respectively.
It can be shown that Im$\chi^f(\omega)$ is related to
$\chi^f_{ii}(\tau)$ by
\begin{equation}
\chi^f_{ii}(\tau)=-\int_{-\infty}^\infty d\omega {{{\mbox{Im}}\chi^f(\omega)
e^{-\tau\omega}} \over {1-e^{-\beta\omega}}},
\label{spindensity}
\end{equation}
with a similar expression relating Im$\Pi^f(\omega)$ to
$\Pi^f_{ii}(\tau)$. In Fig.~4(a) we show -Im$\chi^f(\omega)$ for
several choices of $V$ on a $6\times 6$ lattice. The spin gap
only opens for values of $V$ above the critical point, where there
is no antiferromagnetic long-range order.
In Fig.~4(b) we show -Im$\Pi^f(\omega)$ with the same parameters as in
Fig.~4(a).
In this case it is clear that the charge gap
$\Delta_c$ is present also for values of $V$ below the critical point.
In the inset we report the value of the ratio of the spin gap to the
charge gap $\Delta_s/\Delta_c$ as a function of $V$.
We find that $\Delta_c > \Delta_s$ for all values of
$V$. However, for the larger values of $V$ we can
clearly see that the two gaps tend to be equal. This behavior is an
indication that the system crosses into a band-insulator regime.

Evidence for charge and spin gaps
can also be seen in the temperature dependence of
the uniform charge and spin susceptibilities. The spin
susceptibility is defined by
\begin{equation}
\chi_{\mbox{\small tot}}(q=0)=\int_0^\beta d\tau \langle m(\tau) m(0) \rangle,
\label{bulkspin}
\end{equation}
and the charge susceptibility is similarly defined by
\begin{equation}
\Pi_{\mbox{\small tot}}(q=0)=\int_0^\beta d\tau \left[
\langle n(\tau) n(0) \rangle - \langle n \rangle^2 \right].
\label{bulkcharge}
\end{equation}
Here
$m={1\over N} \sum_i \left( n_{i\uparrow}-n_{i\downarrow}\right)$ and
$n = {1\over N} \sum_i \left( n_{i\uparrow}+n_{i\downarrow}\right)$, with
$n_{i\sigma}=n^f_{i\sigma}+n^d_{i\sigma}$.
In Fig.~5 we show $\chi_{\mbox{\small tot}}$ and $\Pi_{\mbox{\small tot}}$
versus temperature for
several values of the hybridization energy $V$ at a fixed $U_f=4$.
For smaller values of
$V$ we find that $\Pi_{\mbox{\small tot}}$ becomes very small
at a finite temperature, while
$\chi_{\mbox{\small tot}}$
peaks at a temperature close to the one at which we find the
Kondo resonance in $N^f(\omega)$ and then approaches a finite
value at $T=0$. This behavior is consistent with the fact that
for $V=1.0$, Im$\Pi^f(\omega)$ has a gap but Im$\chi^f(\omega)$ is gapless.

In conclusion, we have shown that for small values of $V^2/Ut$
the ground state of the PAM is an insulator
with long-range antiferromagnetic order characterized
by a finite charge gap and gapless spin excitations.
As $V^2/Ut$ increases the long-range order is destroyed
and the system exhibits a spin-liquid behavior. The spin-liquid state
is characterized by both a spin gap and a charge gap with $\Delta_c >
\Delta_s$. When the hybridization $V$ increases further and
$V^2/Ut$ becomes large, the system crosses over to a band-insulating
state in which $\Delta_s$ approaches $\Delta_c$.
Work is currently in
progress to determine the behavior of the doped system.

We thank M.P.A.~Fisher, R.M.~Fye, J.~Freericks, M.~Jarrell, and
A.W.~Sandvik for helpful conversations.
The numerical calculations were performed primarily
on the Cray C-90 at the San Diego Supercomputer Center.
This work was supported by the
National Science Foundation under grants No. DMR92-06023 and
No. PHY89--04035 (Institute for Theoretical Physics) (M.V. and R.T.S.), the
Los Alamos National Laboratory under the  LACOR grant No.
UC-94-2-A-213 (M.V.), and
the Department of Energy under grant No. DE-FG03-85ER45197 (D.J.S. and
R.L.S.).

\begin{figure} FIG. 1.
The square of the staggered magnetization $m^2_f$ for various
$\sqrt{N}\times \sqrt{N}$
lattices as a function of $1/\sqrt{N}$. The solid
symbols are extracted using  Eq.(\ref{spinwave1}) and the open symbols
using Eq.(\ref{spinwave2}). The circles are for $V=1.0$ and the triangles
for $V=1.2$. Here $U_f=4$ and $\beta=20$. For comparison, the solid
square indicates the value of $m_f^2$ for the 2D Hubbard model with $U=4$.
\end{figure}
\begin{figure} FIG. 2.
The $U_f$-$V$ phase diagram of the half-filled PAM at $T=0$, showing the
boundary between the antiferromagnetic (AF) and spin-liquid (SL) phases.
The solid line is a guide to the eye through the QMC points.
The dotted line is the MFT boundary.
\end{figure}
\begin{figure} FIG. 3.
The $f$-band density of states $N^f(\omega)$ on a $6\times 6$ lattice with
$U_f=4$ and $V=0.75$ for several temperatures. The dashed line is for
$\beta=4$, the dashed-dotted line for $\beta=12$ and the solid line for
$\beta=20$.
\end{figure}
\begin{figure} FIG. 4.
The imaginary part of (a) the spin susceptibility -Im$\chi^f(\omega)$
and (b) the charge susceptibility -Im$\Pi^f(\omega)$ for $\beta=16$ on
a $6\times 6$ lattice with $U_f=4$.
The ratio of the spin and charge
gaps $\Delta_s/\Delta_c$ versus $V$ is shown in the inset.
\end{figure}
\begin{figure} FIG. 5.
The uniform (a) spin susceptibility $\chi_{\mbox{\small tot}}(q=0)$ and
(b) charge susceptibility $\Pi_{\mbox{\small tot}}(q=0)$ as a function of
temperature $T$ on a $6\times 6$ lattice with $U_f=4$.
\end{figure}

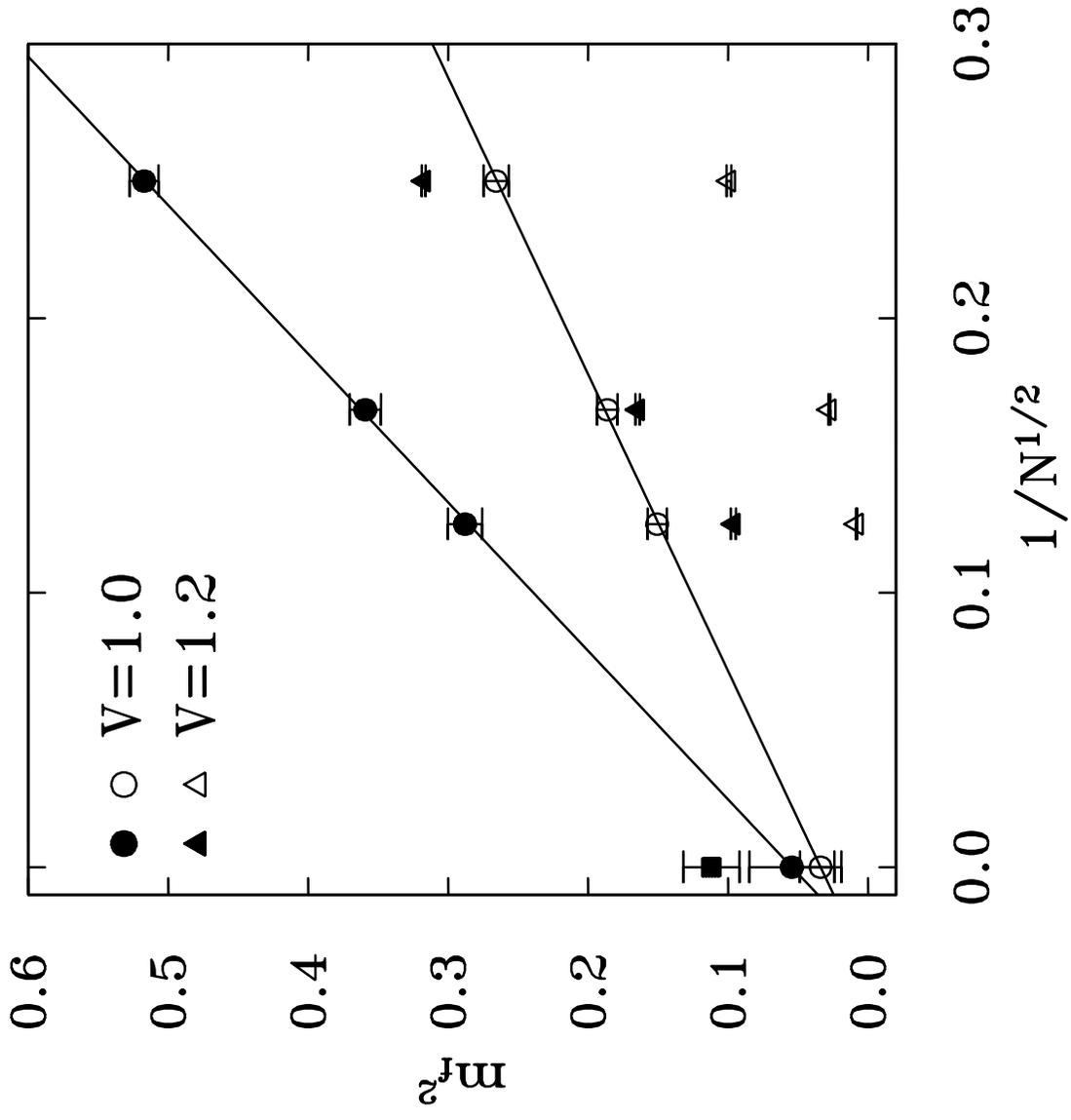
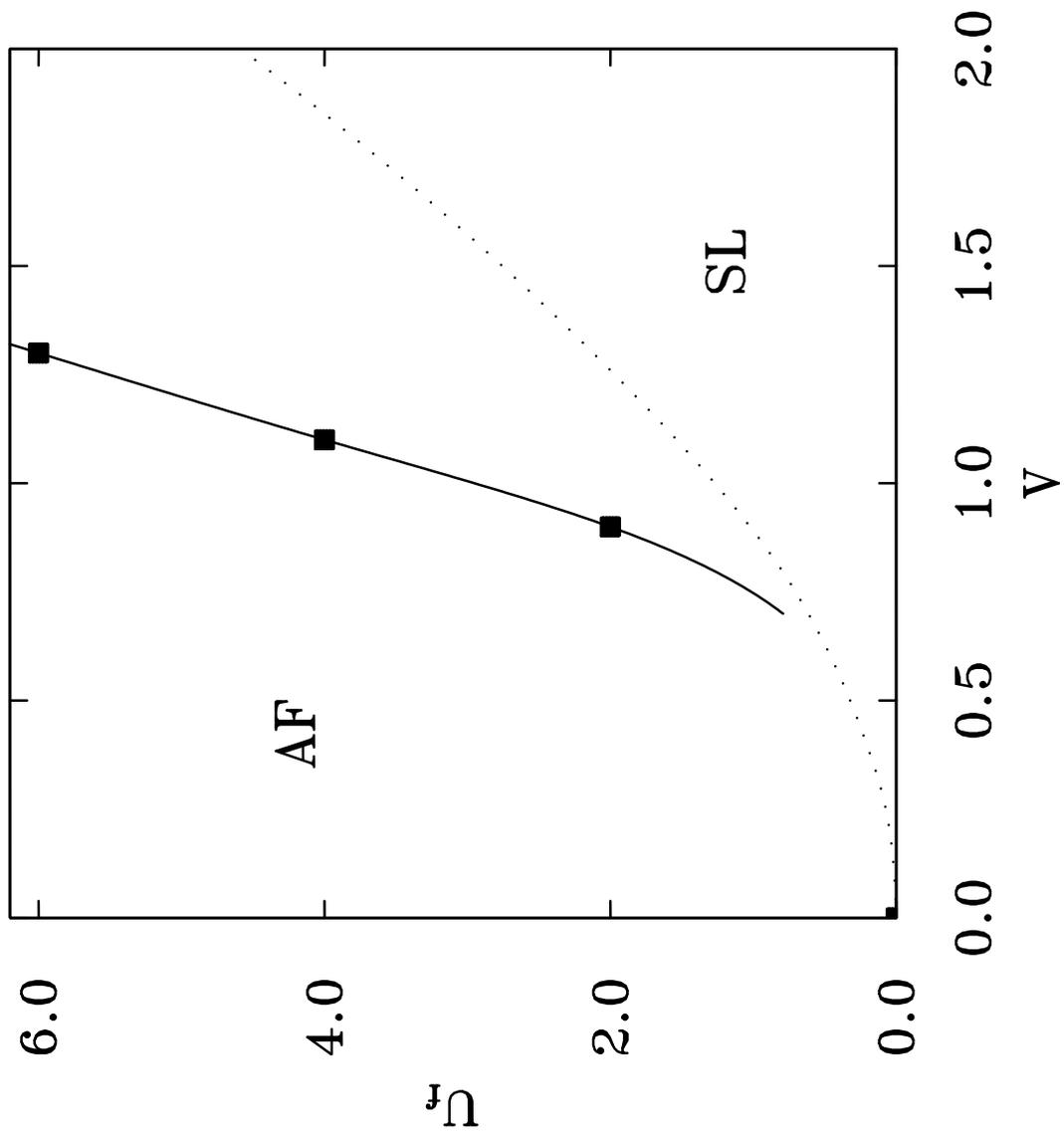
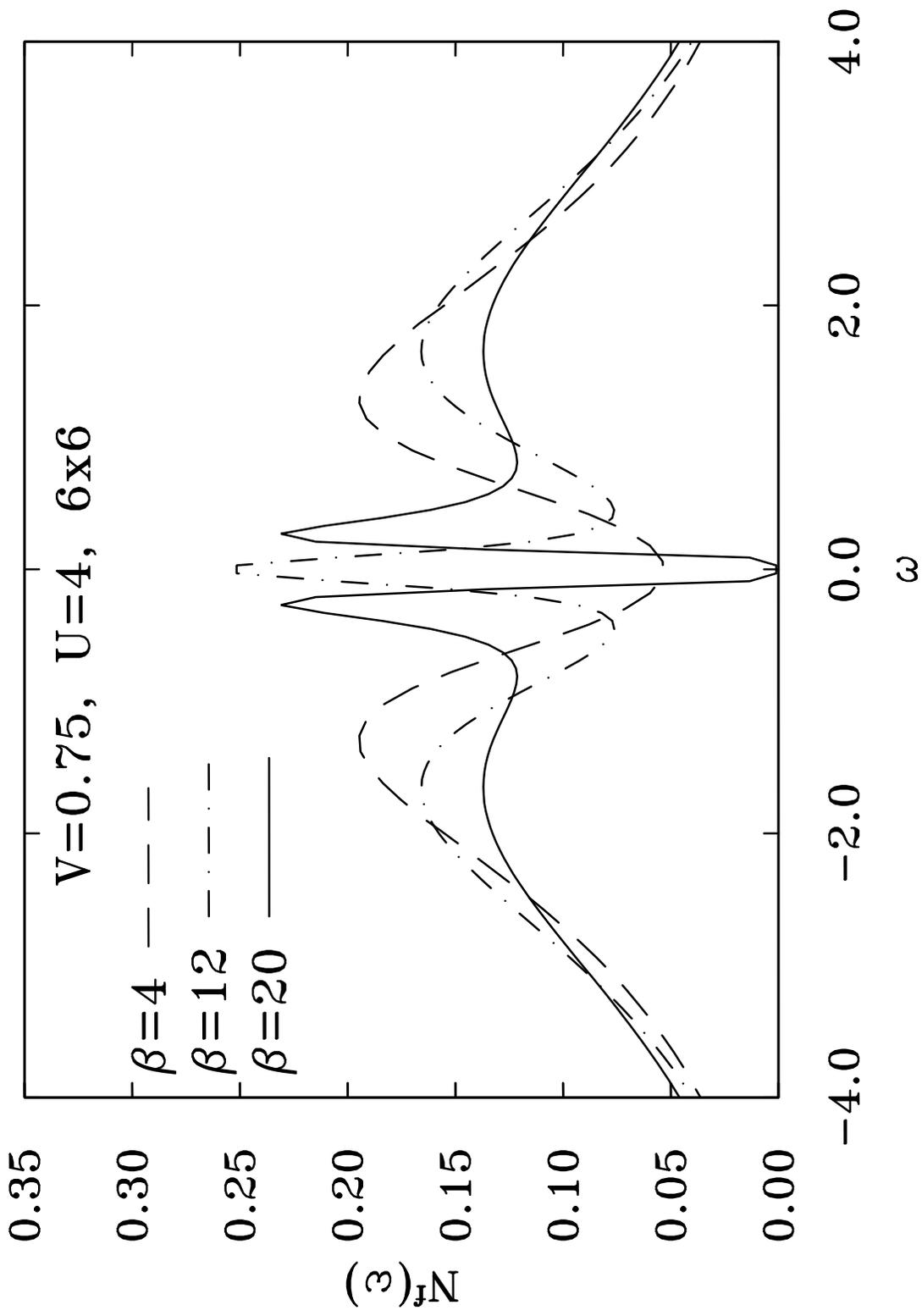
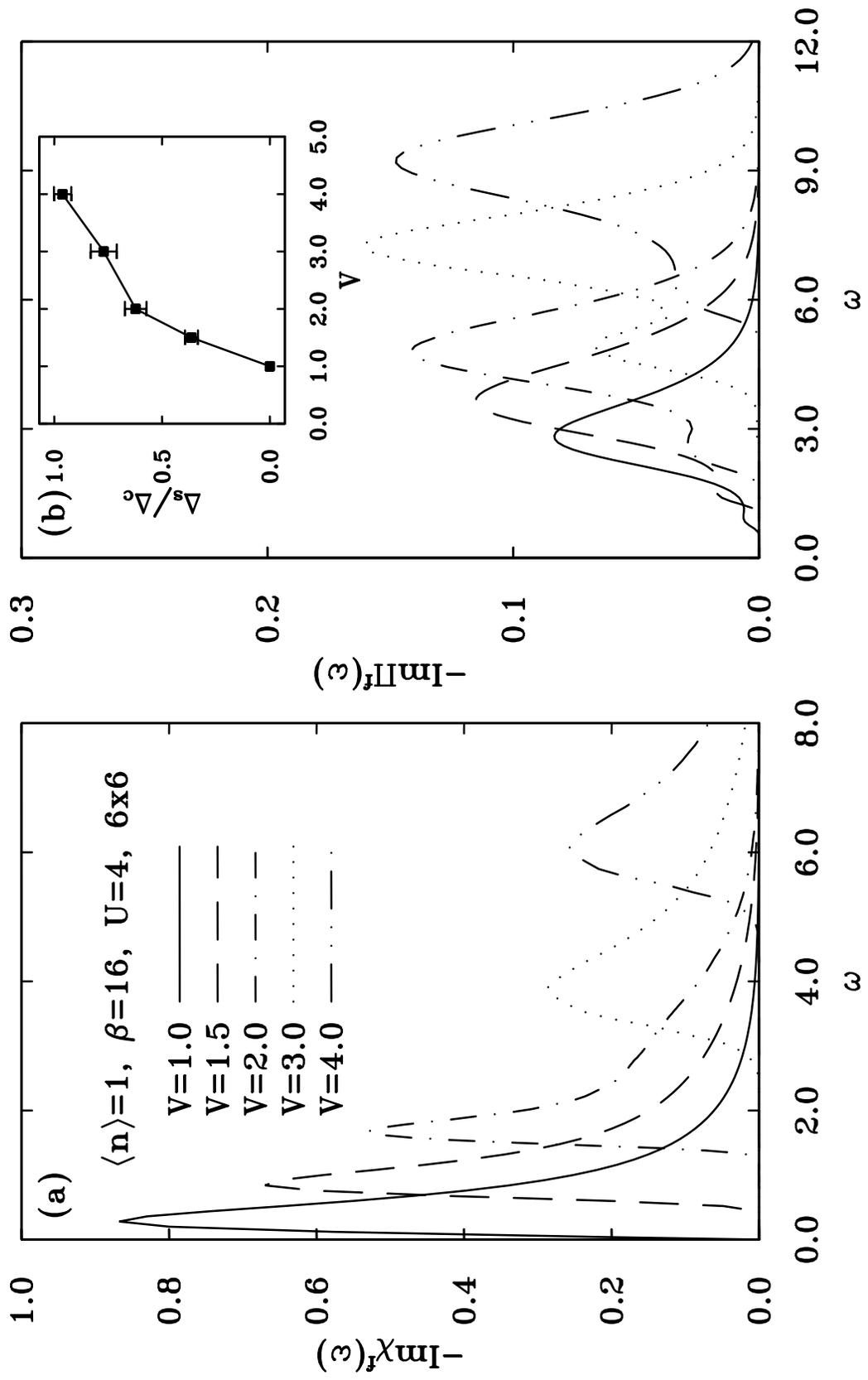
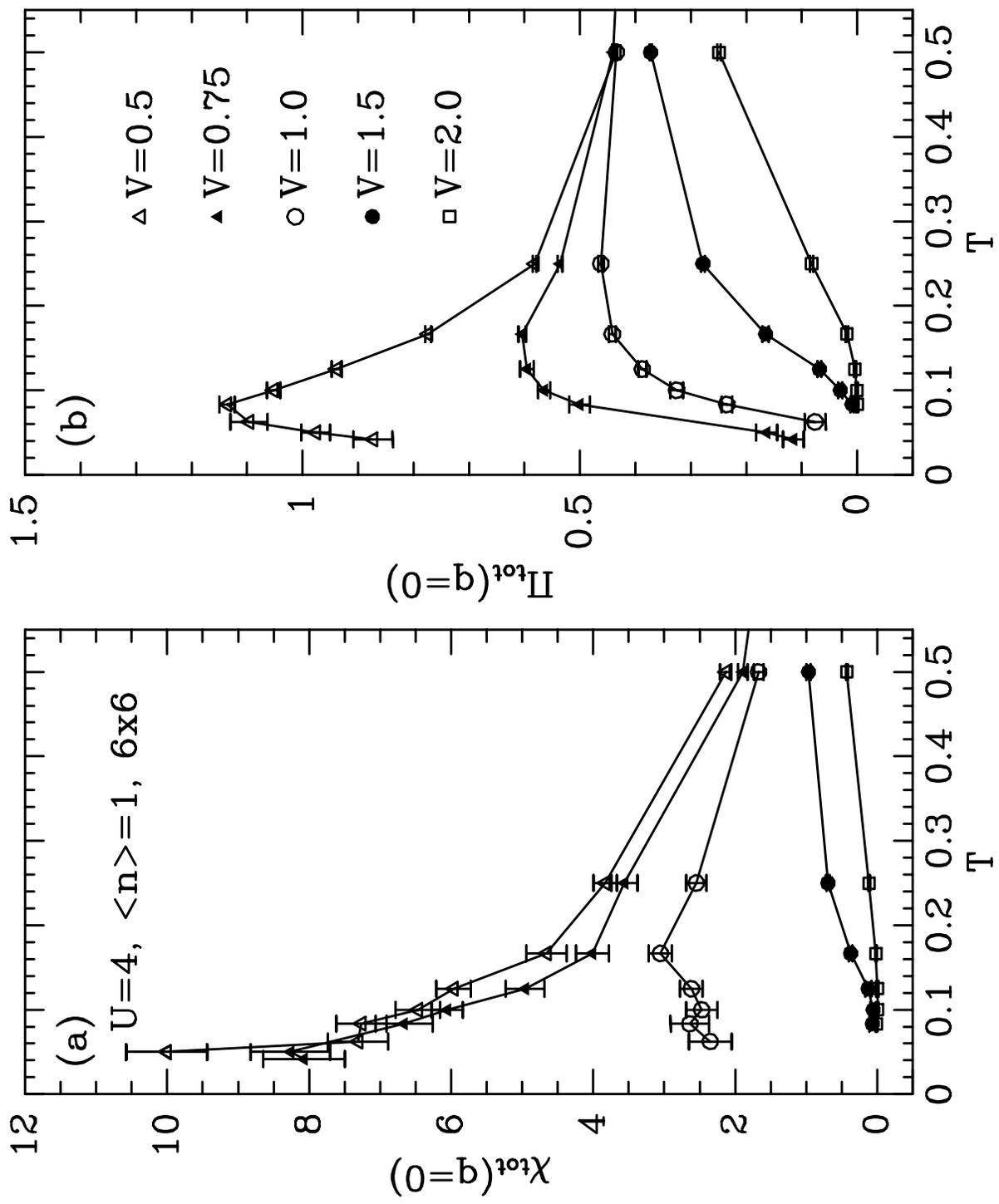
\end{document}